\documentclass{article}
 \usepackage{amsmath}
 \usepackage{bm}
 \usepackage{graphicx}
\begin{document}

%\preprint{}

\title{Multiple Ultrasound Image Generation based on Tuned Alignment of Amplitude Hologram over Spatially non-Uniform Ultrasound Source}%Title of paper
 \author{Keisuke Hasegawa}
 % \correspondingauthor
\maketitle

\section*{abstract}
In this study, a method for readily and inexpensively generating real-time reconfigurable intense midair ultrasound field is proposed.
Recent investigations and applications of midair convergent high-power ultrasound have been increasingly growing.
For generating such ultrasound fields, specifically designed ultrasound sources or phased arrays of ultrasound transducers are conventionally used.
The former can be more readily fabricated but cannot drastically reconfigure the generated ultrasound field, and the latter can create electronically controllable ultrasound fields but is much more difficult to implement and expensive.
The proposed method utilizes a planar ultrasound source with a fixed surface vibration pattern and a newly designed amplitude mask inspired by the Fresnel-zone-plate (FZP).
The mask is designed so that when it is placed on a specific position on the source, the partially covered source emission results in forming pre-determined ultrasound fields with corresponding specific spatial pattern.
With these new masks, the generated fields can be switched among several presets by changing the mask position on the source.
The proposed technique only requires slight mechanical translation of the mask over the source to instantaneously reconfigure the resulting midair ultrasound field.
The proposed method enables one to create a reconfigurable ultrasound field with a large source aperture in a practically feasible setup, which will potentially broaden the workspace of current midair-ultrasound applications to the whole-room scale.
\footnote{K. H. Author is to whom corresponding should be addressed (e-mail: keisuke@mech.saitama-u.ac.jp).\\
Department of Mechanical Science, Graduate School of Science and Engineering, Saitama Universlty, Saitama-city, 338-8570, Japan.\\
This project is financially supported by Japan Society for the Promotion of Science (JSPS), Grant No. 23K18488, Japan.}

 %\maketitle must follow title, authors, abstract and \pacs

Many applications using nonlinear acoustic effects of midair convergent ultrasound have been developed.
Specifically, its non-contact nature that can be utilized in mechanical manipulation is a significant advantage unique to this effect.
Acoustic radiation force \cite{King1934, Hamilton1998, Sapozhnikov2013} and acoustic streaming \cite{Eckart1948, Nyborg1953} are the representative nonlinear effects, which are utilized in acoustic levitation \cite{Xie2002, Hirayama2019, Morales2022}, non-contact mechanical sensing \cite{Fujiwara2011}, non-contact haptic stimulation \cite{Hoshi2010, Takahashi2020, Nakajima2021, Frier2018}, midair transportation of gaseous materials \cite{Hasegawa2018, Hasegawa2019}, etc. 
For those applications, it is necessary to converge acoustic energy within appropriately small regions in a manner that their spatial distribution matches the desired one for specific applications.

To realize such intense ultrasound field in air, an ultrasound emission aperture larger than the wavelength is widely used.
In general, emission patterns on the aperture is created non-uniform; the phase and amplitude on the emission plane may spatially vary and thus result in spatially localized midair ultrasound fields.
The designing method of emission pattern for realizing a given ultrasound field has been investigated, which is known as acoustic holography calculation techniques \cite{Gerchberg1972, Mellin2001, Leonardo2007}.
\begin{figure*}
%\vspace{-0.3in}
\includegraphics[width=4.7in]{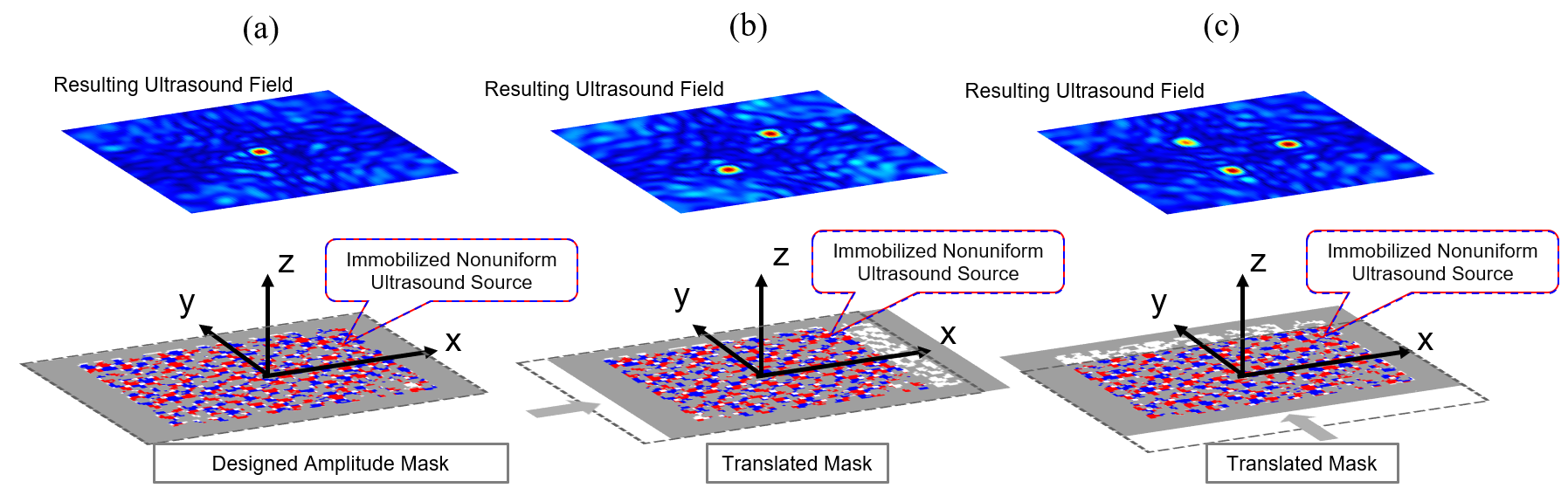}
\caption{Schematic description of the concept of selective multiple ultrasound field generation with a translated amplitude mask on a non-uniform vibrating source. The generated ultrasound fields drastically change corresponding to small spatial shifts of the mask (a)-(c).}
\label{fig1}
\end{figure*}

For hardware construction, there are two categories of methods for practically generating such non-uniform ultrasound emission apertures.
One is physically constructing an emission plane with a designed spatial distribution of phase delays and amplitudes.
This includes emission plane with uneven height \cite{Melde2016, Poly2020}, acoustic element arrays for the phase delay function \cite{Jahdali2016, Zhao2018}, use of metamaterials \cite{Hennion2013, Zhao2020}.
Some of these methods can be relatively inexpensive and readily implemented.
However, once fabricated, resulting midair ultrasound fields by them cannot be altered except for a few configurable ones \cite{Memoli2017, Zhao2018}, which requires considerable time for actual reconfiguration.
Another major method is the use of phased arrays of midair ultrasound transducers \cite{autd3, Marzo2019}. By appropriately controlling the driving amplitudes and phase delays of individual transducers, the ultrasound field with the desired spatial distribution can be generated in an electronically controllable manner.
Due to this advantage, they are applied to applications that require dynamic ultrasound field control \cite{Morales2019, Hirayama2022}.
On the other hand, the construction of the phased arrays with a large number of transducer elements is costly and time-consuming, which prevents the prevalent use of this technique in a practical context.

In this study, I focus on the use of an amplitude mask over the ultrasound source, inspired by the Fresnel zone plate (FZP), which is categorized into the former methods.
Usually, an FZP refers to an amplitude mask plate on which concentric aperture rings that transmit ultrasound are made.
Incident planer waves are partially blocked off by the FZP and converted into a cluster of spherical waves on the aperture rings, resulting in converging at a focusing point whose depth can be tuned by the intervals of the aperture rings.
There have been many FZP-inspired methods proposed for midair ultrasound focusing \cite{Schindel1997,Xia2020, Perez2021}, including ours demonstrating that a horizontal shift of an FZP placed on the ultrasound transducer array driven in-phase results in accordingly shifted position of the ultrasound focus \cite{Kitano2023}.
This method is readily used for ultrasound focusing, but drastic change and instantaneous in the generated midair ultrasound field still cannot be achieved.

In this paper, a new strategy is proposed to generate mechanically and instantaneously reconfigurable ultrasound field using the following two components: an ultrasound emitting source with nonuniform phase distribution and an amplitude mask newly designed to block off specific emission region of the source according to its relative position with respect to the source. 
Unlike most of the FZP-based focusing methods, the proposed method does not assume the incident wave to be planar.
The phase distribution of the source (the incident wave) is supposed to be known in prior, and the relative position of the mask on it determines which parts of the nonuniform emission is blocked off.
Therefore, the resulting ultrasound field is entirely different depending on the mask position, unlike the case with plane wave emission.
In the proposed method, the mask is designed so that when placed on specific positions on the source, the partially covered source emission results in forming pre-determined ultrasound field with corresponding specific spatial patterns.

The idea of converging spatially nonuniform emission into pinpoint regions using an amplitude mask are in common with our previous work \cite{Hasegawa2025}.
What differentiates this study from the previous one is that multiple preset emission patterns are recorded in the mask which corresponds to horizontal shift of the mask on the source by several millimeters (Fig. \ref{fig1}.)
In other words, slight mechanical translation of the plate results in drastic change in generated ultrasound field.
Such small displacement of the plate is readily and inexpensively implemented, and thus the proposed method offers virtually instantaneous reconfiguration of the generated intense ultrasound field with a very simple configuration, which has not been realized by conventional ultrasound focusing methods free of use of phased arrays. 

In the following part of the paper, the designing procedure of the amplitude mask is described, as well as the numerical simulations and measurement results of the generated ultrasound field.
It should be noted that the ultrasound source used in the experiment is the phased array for experimental convenience.
In practice, an array of ultrasound transducers can be used instead, whose binary driving phase delays are predetermined by the electrical circuitry.
This is much more easily implemented because only a pair of amplifiers are required that provide driving current to individual transducers via proper wiring according to the transducer position on the emission plane.
Both emission planes of transducers and masks can be inexpensively fabricated with a large spatial dimension.
Therefore, the proposed method has a great potential in realizing a large reconfigurable emission plane of midair ultrasound field.
Such emission planes can extend current midair ultrasound systems to the whole room scale and a scope of conceivable application scenario will be tremendously widened.

\begin{figure*}
%\vspace{-0.3in}
\includegraphics[width=4.7in]{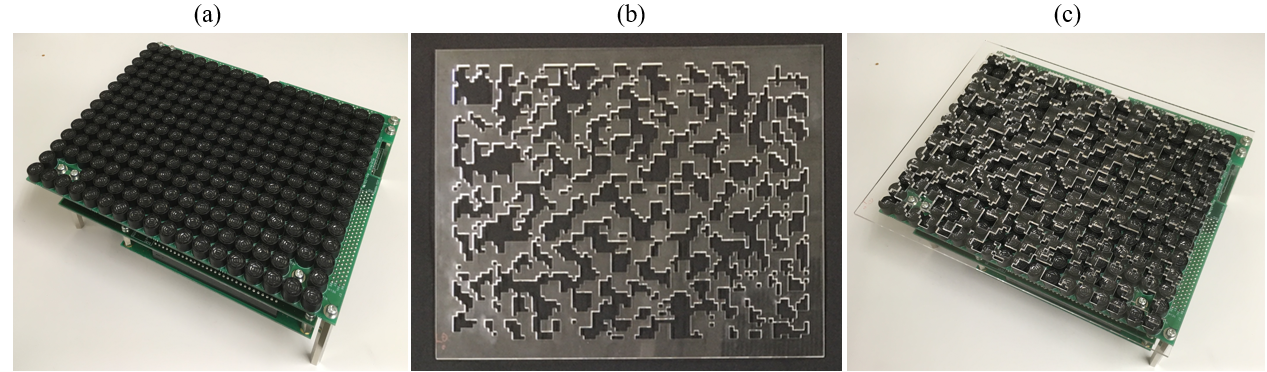}
\caption{(a) Ultrasound phased array used as an emission source, (b) fabricated amplitude mask, and (c) the mask placed on the emission source as in the experiments. }
\label{fig2}
\end{figure*}

%Methods
Suppose that the emission plane of the source is composed of a finite number of omnidirectional point sources.
The transparency pattern of the mask is often designed via the phase-amplitude conversion principle, where the amplitude pattern of emission is calculated by thresholding the phase pattern of emission that corresponds to the desired ultrasound field.
The phase pattern on the emission plane for the desired resulting ultrasound fields can be obtained via several methods like the time-reversal methods \cite{Fink1992}, for example.

This process is understood as approximating the spatial phase distribution of the emission plane as binary, and the ‘negative' phase parts of the emission are blocked off by the mask so that the rest of the ‘positive’ phase parts of the emission constructively interfere in the desired region in the air.

Conventionally, the mask is designed to create a single fixed ultrasound field out of partial occlusion of planar incident wave.
In this study, the mask is designed to selectively create an ultrasound field out of multiple presets, by calculated partial occlusion of ultrasound emission with nonuniform phase distribution.
Prior to the designing of the mask, calculation of the corresponding phase distribution patterns of emission plane that gives the desired ultrasound fields is followed by.

Suppose that $M$ different ultrasound fields are to be created, and $p_j(\bm{r}), j = 1,\ldots, M$ denotes the corresponding emission patterns in complex amplitude on $\bm{r} = (x, y, z)$ when the mask is placed on the source on the $j$-th relative position.
Similarly, let $\bm{d}_j$ be the corresponding translation of the mask over the source.
Here, let $s(\bm{r})$ be the complex amplitude distribution of the source, which is equal to 1 for where its driving phase is $0$ and is equal to 1 for where its driving phase is $\pi$.
Then, the amplitude-phase pattern $m_c(\bm{r})$ on the emission plane comprising complex amplitude on $\bm{r}$ is desired to hold $p_j(\bm{r}) = m_c(\bm{r} - \bm{d}_j)s(\bm{r})$ for $j = 1,\ldots, M$.
This is equivalent to have $p'_j(\bm{r}) = m_c(\bm{r})s_j(\bm{r})$, where $p'_j(\bm{r}) = p_j(\bm{r} + \bm{d}_j), s_j(\bm{r}) = s(\bm{r} +\bm{d}_j)$.
However, the number of the constraints are $M$ times larger than that of the tunable parameters $m_c(\bm{r})$ for individual $\bm{r}$, so the minimum-total-squared-error criteria is introduced here to determine optimal values of $m_c(\bm{r})$. 

By minimizing the total squared error given as
\begin{equation}
\sum_{j = 1}^M \left|p'_j(\bm{r}) - m_c(\bm{r})s_j(\bm{r})\right|^2,
\end{equation}
the optimal value of $m_c(\bm{r})$ is obtained as
\begin{equation}
m_c(\bm{r}) = \sum_{j = 1}^M \frac{s_j^*(\bm{r})p'_j(\bm{r})}{s_j^*(\bm{r})s_j(\bm{r})}.
\end{equation}
Note that the all variables in the above equations are complex number and $|\cdot|$ and $\cdot^*$ denote the absolute value of and the complex conjugate of $\cdot$, respectively.
As conventionally done, the calculated complex amplitude distribution on the emission plane $m_c(\bm{r})$ is converted into binary amplitude mask $m(\bm{r})$ in line with the following rule:
\begin{align}
m(\bm{r}) = \left\{
\begin{array}{cc}
1, & 0 \leq \angle m_c(\bm{r}) < \pi\\
0, & -\pi \leq \angle m_c(\bm{r}) < 0,
\end{array} 
\right.
\end{align}
where $\angle \cdot$ denotes the argument of $\cdot$.
In the calculation of $m(\bm{r})$, the denominator of Eq. (2) can be zero for some $\bm{r}$, which happens when $s_j(\bm{r}) = 0$ for all $j$.
In this case, no sound emission is observed at such $\bm{r}$ for all $M$ source shifts. 
Therefore, the transparency amplitude on the mask $m(\bm{r})$ at such a point can be either 0 or 1.
In this study, $m(\bm{r})$ is set to 0 for the case.

In this study, a custom-made airborne ultrasound phased array \cite{autd3} was used as an ultrasound source with its driving phase distribution fixed.
Therefore, no phase engineering was performed during the experiments.
The driving frequency of the transducer was 40~kHz, resulting in the wavelength of 8.5~mm in the air for the sound velocity of 340~m/s.
The phased array contains cylindrical ultrasound transducers aligned in a two-dimensional lattice of 18 columns and 14 rows.
The phased array was driven so that the phase distributions on the transducers yielded a checker-pattern, meaning that every neighboring transducer had opposite driving phase of either 0 or $\pi$.
The coordinate system of the experiment is depicted in Fig. 1.
The desired emission patterns $p_j(\bm{r})$ was set for $j = 1,2,3$: $p_1(\bm{r})$ corresponds to a single ultrasound focus placed at $(0\mathrm{~mm},0\mathrm{~mm},180\mathrm{~mm})$, $p_2(\bm{r})$ corresponds to two ultrasound foci at $(30\mathrm{~mm},30\mathrm{~mm},180\mathrm{~mm})$ and $(-30\mathrm{~mm},-30\mathrm{~mm},180\mathrm{~mm})$, and $p_3(\bm{r})$ corresponds to three ultrasound foci at $(40\mathrm{~mm},0\mathrm{~mm},180\mathrm{~mm})$, $(-20\mathrm{~mm},34.6\mathrm{~mm},180\mathrm{~mm})$, and $(20\mathrm{~mm},34.6\mathrm{~mm},180\mathrm{~mm})$.

The distribution of $p_j(\bm{r})$ was calculated as summation of the complex amplitude pattern for generating individual focus.
The phase delay compensation of each transducer was given according to the distance from the focal point:
\begin{align}
p_j(\bm{r}) = \sum_{n=1}^{N_j}e^{ik||\bm{r} - \bm{f}_{jn}||},   
\end{align}

where $N_j$ is the numbers of foci and $\bm{f}_{jn}$ is the $n$-th focal position in $j$-th emission patterns, $k$ is the wavenumber, and $i = \sqrt{-1}$ is the imaginary unit.
The translation of the mask was set to $\bm{d}_1 = (0\mathrm{~mm},0\mathrm{~mm},0\mathrm{~mm}), \bm{d}_2 = (6\mathrm{~mm},0\mathrm{~mm},0\mathrm{~mm})$ and $\bm{d}_3 = (0\mathrm{~mm},6\mathrm{~mm},0\mathrm{~mm})$. 
\begin{figure*}
%\vspace{-0.3in}
\includegraphics[width=4.7in]{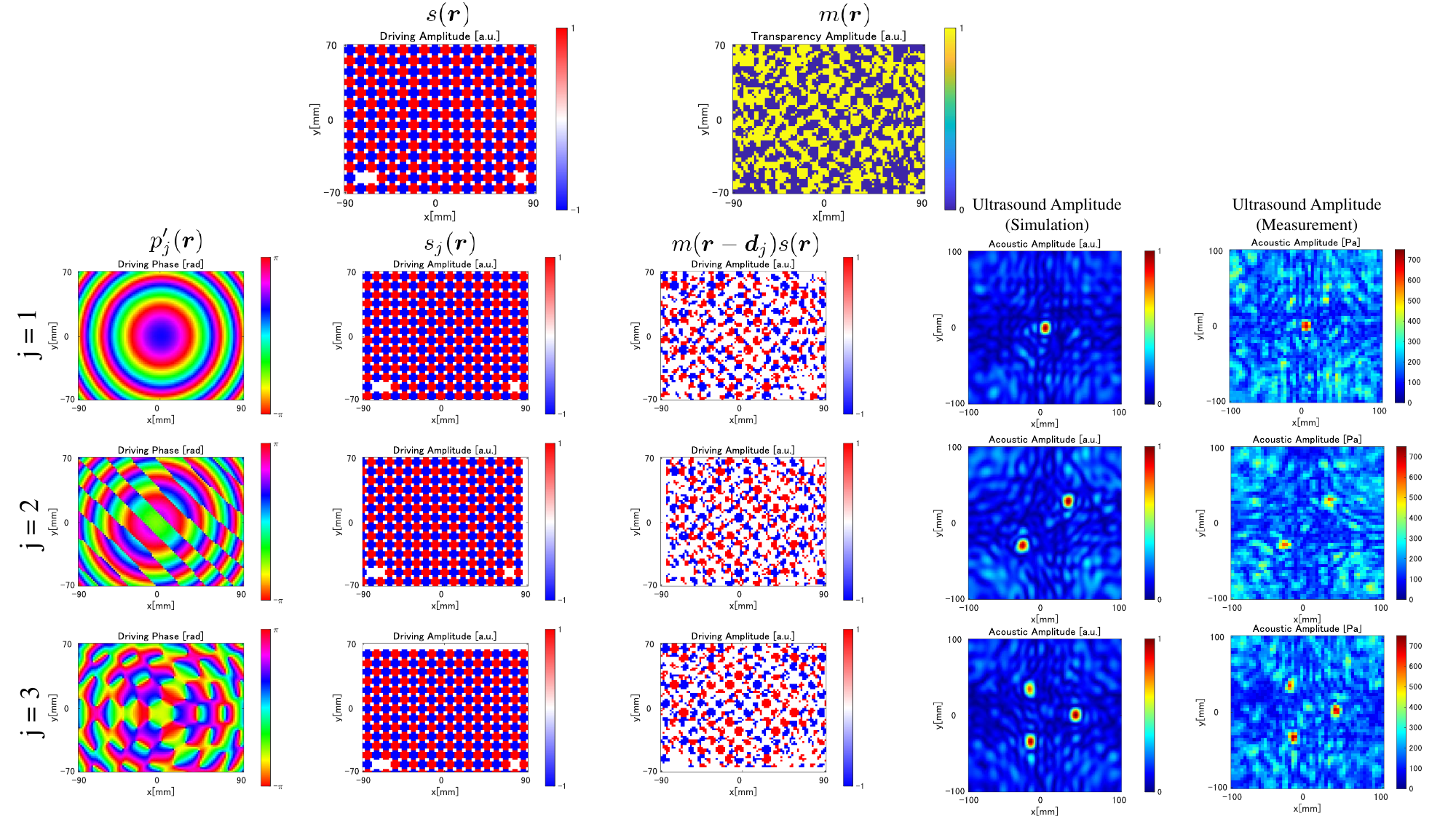}
\caption{The driving amplitude distribution on the source plane and the designed masking pattern in the experiments (Upper figures), and the distribution of $p_j'(\bm{r}), s_j(\bm{r}), m(\bm{r} - \bm{d}_j)s(\bm{r})$, numerically and experimentally obtained ultrasound fields on the focal plane, for the given three patterns in the experiments ($j = 1,2,3.)$ }
\label{fig3}
\end{figure*}

Figure 2 shows the phased array used in the experiment and the mask created by cutting out the acrylic plate with a thickness of 2~mm using a laser cutting machine.
In line with the calculation results in Figure 3, the cutting pattern was designed as a cluster of 2~mm-length squares whose positions corresponded to where $m(\bm{r}) = 1$, which let the emitted ultrasound go through.
When two or more squares are adjacent, they coalesced to form a larger ultrasound aperture.
As a result, some regions that should block off the ultrasound were left isolated from the edges of the plate.
For those regions, some neighbouring squares were manually deleted to prevent them from falling after fabricated, by connecting them to the edges. 

The distribution of $p'_j(\bm{r}), s_j(\bm{r})$ and the calculated amplitude mask $m(\bm{r})$ are displayed in Fig. 3, along with the masked emission patterns and resulting ultrasound amplitude fields obtained via numerical calculation and physical measurement.
In the numerical ultrasound field calculation, the masked sound sources are modeled as a cluster of point sources inside the transducer surfaces and the discretization interval of the source emission and the resulting field was all set to 2~mm.
The physical measurement of acoustic scanning was performed using two-dimensional linear actuators (EZSM6 and EZSM4 series, products of Oriental Motor, Japan) on which a standard microphone system (1/8-in. microphone, type 4138-A-015; pre-amplifier, type 2670; condition amplifier, type 2690-A; all products of Hottinger, Br\"{u}el and Kj\ae r GmbH, Germany) was mounted.
The scanning interval was 4~mm.
In both numerical simulation and the measurement, distance between the phased array surface and the planer region of interest was set to 180~mm. 

\begin{figure*}
%\vspace{-0.3in}
\includegraphics[width=4.7in]{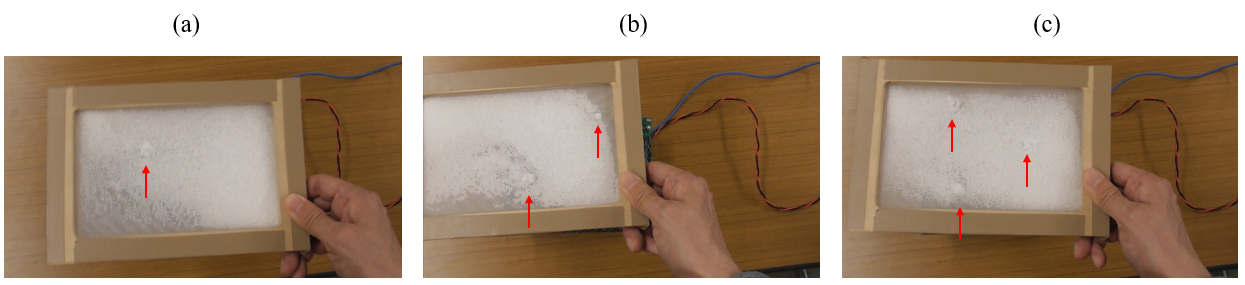}
\caption{The resulting ultrasound fields are visualized via polystyrene beads confined between acoustically translucent layers being pushed upward \cite{Hasegawa2023} with the amplitude mask correspondingly translated. Three figures correspond to the three patterns with (a) single ($j = 1$), (b) double ($j = 2$), and (c) triple ($j = 3$) focusing.} %Full video is available in the Supplementary Materials.} 
\label{fig4}
\end{figure*}

The numerical and experimental results of ultrasound field generation show good agreement, indicating that the proposed method can successfully realize reconfigurable ultrasound field generation out of slight mechanical shift of the mask over sound sources.
The generated ultrasound fields included ultrasound foci with desired number and positions, which were strong enough to cause nonlinear acoustic effects such as acoustic radiation pressure, for example (visualized via polystyrene beads confined between acoustically translucent layers \cite{Hasegawa2023} in Fig. 4.)
At the same time, fairly intense artifacts scattering around the scattering focal patterns can be seen especially in the measurement results, which is the drawback of this binary amplitude emission control method.
Such scattering artifacts are much less prominent in cases with phase-controlling method of emissions \cite{Kitano2023}.

The individual emission pattern in this study is calculated as summation of complex amplitude patterns on the sources that cancel the phase delay to form each focus.
Also, the amplitude pattern design of the mask is completed via simple total squared error minimization scheme, and the phase-to-amplitude conversion procedure does not take the amplitude of $m_c(\bm{r})$ into consideration.
In addition, the source pattern $s(\bm{r})$ is supposed to be given in this study, which may also be designed according to the ultrasound field patterns to be generated.
Those technical aspects may be amended to improve the performance of the proposed amplitude-mask-based ultrasound field generation methods, hopefully by suppressing the artifact, for example.

In conclusion, a method of generating reconfigurable intense midair ultrasound fields out of nonuniform emission of planar ultrasound sources partially blocked off by the newly designed amplitude mask was proposed, where its slight translational shift completely switches the generated field pattern.
The designing of the amplitude mask can be completed with simple squared error minimization calculation and fabrication of the mask is readily done by laser cutting machine using inexpensive materials.
The airborne ultrasound phased array with fixed phase distribution is used in this study as an emission source, but it can be replaced with large monolithic vibrating plate with its vibration mode theoretically or experimentally known.
Then, it is expected to be possible to use such plates as emission sources and generate configurable ultrasound field in the same way as this study by designing the mask pattern according to their vibration modes.
Using the methods of this study, the midair ultrasound application based on reconfigurable intense ultrasonic emission would no longer be the privileges of specialists who possess fairly expensive ultrasound phased arrays.
Thus, it is expected that an increased number of users who can afford to utilize the proposed method will contribute to invention of new applications in this field.

This project is financially supported by JSPS, Grant No. 23K18488, Japan.

\section*{AUTHOR DECLARATIONS}
\subsection*{Conflict of Interest}
The authors have no conflicts to disclose.
\subsection*{Author Contributions}
K. Hasegawa: Conceptualization; Data curation; Formal analysis; Investigation; Methodology;
Visualization; Writing; Funding acquisition; Project administration; Validation.
\subsection*{Data Availability}
The data that support the findings of this study are available from
the corresponding author upon reasonable request.

\bibliographystyle{amsplain}
\bibliography{aipsamp}

\providecommand{\noopsort}[1]{}\providecommand{\singleletter}[1]{#1}%
\begin{thebibliography}{36}%
\makeatletter
\providecommand \@ifxundefined [1]{%
 \@ifx{#1\undefined}
}%
\providecommand \@ifnum [1]{%
 \ifnum #1\expandafter \@firstoftwo
 \else \expandafter \@secondoftwo
 \fi
}%
\providecommand \@ifx [1]{%
 \ifx #1\expandafter \@firstoftwo
 \else \expandafter \@secondoftwo
 \fi
}%
\providecommand \natexlab [1]{#1}%
\providecommand \enquote  [1]{``#1''}%
\providecommand \bibnamefont  [1]{#1}%
\providecommand \bibfnamefont [1]{#1}%
\providecommand \citenamefont [1]{#1}%
\providecommand \href@noop [0]{\@secondoftwo}%
\providecommand \href [0]{\begingroup \@sanitize@url \@href}%
\providecommand \@href[1]{\@@startlink{#1}\@@href}%
\providecommand \@@href[1]{\endgroup#1\@@endlink}%
\providecommand \@sanitize@url [0]{\catcode `\\12\catcode `\$12\catcode
  `\&12\catcode `\#12\catcode `\^12\catcode `\_12\catcode `\%12\relax}%
\providecommand \@@startlink[1]{}%
\providecommand \@@endlink[0]{}%
\providecommand \url  [0]{\begingroup\@sanitize@url \@url }%
\providecommand \@url [1]{\endgroup\@href {#1}{\urlprefix }}%
\providecommand \urlprefix  [0]{URL }%
\providecommand \Eprint [0]{\href }%
\providecommand \doibase [0]{http://dx.doi.org/}%
\providecommand \selectlanguage [0]{\@gobble}%
\providecommand \bibinfo  [0]{\@secondoftwo}%
\providecommand \bibfield  [0]{\@secondoftwo}%
\providecommand \translation [1]{[#1]}%
\providecommand \BibitemOpen [0]{}%
\providecommand \bibitemStop [0]{}%
\providecommand \bibitemNoStop [0]{.\EOS\space}%
\providecommand \EOS [0]{\spacefactor3000\relax}%
\providecommand \BibitemShut  [1]{\csname bibitem#1\endcsname}%
\let\auto@bib@innerbib\@empty
%</preamble>
\bibitem [{\citenamefont {King}(1934)}]{King1934}%
  \BibitemOpen
  \bibfield  {author} {\bibinfo {author} {\bibfnamefont {L.~V.}\ \bibnamefont
  {King}},\ }\bibfield  {title} {\enquote {\bibinfo {title} {On the acoustic
  radiation pressure on spheres},}\ }\href {\doibase 10.1098/rspa.1934.0215}
  {\bibfield  {journal} {\bibinfo  {journal} {Proceedings of the Royal Society
  of London. Series A - Mathematical and Physical Sciences}\ }\textbf {\bibinfo
  {volume} {147}},\ \bibinfo {pages} {212--240} (\bibinfo {year}
  {1934})}\BibitemShut {NoStop}%
\bibitem [{\citenamefont {Wang}\ and\ \citenamefont
  {Lee}(1999)}]{Hamilton1998}%
  \BibitemOpen
  \bibfield  {author} {\bibinfo {author} {\bibfnamefont {T.~G.}\ \bibnamefont
  {Wang}}\ and\ \bibinfo {author} {\bibfnamefont {C.~P.}\ \bibnamefont {Lee}},\
  }\enquote {\bibinfo {title} {Nonlinear acoustics (edited by m. f. hamilton
  and d. t. blackstock)},}\ \ (\bibinfo  {publisher} {Academic Press},\
  \bibinfo {year} {1999})\ Chap.~\bibinfo {chapter} {6}\BibitemShut {NoStop}%
\bibitem [{\citenamefont {Sapozhnikov}\ and\ \citenamefont
  {Bailey}(2013)}]{Sapozhnikov2013}%
  \BibitemOpen
  \bibfield  {author} {\bibinfo {author} {\bibfnamefont {O.~A.}\ \bibnamefont
  {Sapozhnikov}}\ and\ \bibinfo {author} {\bibfnamefont {M.~R.}\ \bibnamefont
  {Bailey}},\ }\bibfield  {title} {\enquote {\bibinfo {title} {{Radiation force
  of an arbitrary acoustic beam on an elastic sphere in a fluid}},}\
  }\href@noop {} {\bibfield  {journal} {\bibinfo  {journal} {The Journal of the
  Acoustical Society of America}\ }\textbf {\bibinfo {volume} {133}},\ \bibinfo
  {pages} {661--676} (\bibinfo {year} {2013})}\BibitemShut {NoStop}%
\bibitem [{\citenamefont {Eckart}(1948)}]{Eckart1948}%
  \BibitemOpen
  \bibfield  {author} {\bibinfo {author} {\bibfnamefont {C.}~\bibnamefont
  {Eckart}},\ }\bibfield  {title} {\enquote {\bibinfo {title} {Vortices and
  streams caused by sound waves},}\ }\href {\doibase 10.1103/PhysRev.73.68}
  {\bibfield  {journal} {\bibinfo  {journal} {Phys. Rev.}\ }\textbf {\bibinfo
  {volume} {73}},\ \bibinfo {pages} {68--76} (\bibinfo {year}
  {1948})}\BibitemShut {NoStop}%
\bibitem [{\citenamefont {Nyborg}(2005)}]{Nyborg1953}%
  \BibitemOpen
  \bibfield  {author} {\bibinfo {author} {\bibfnamefont {W.~L.}\ \bibnamefont
  {Nyborg}},\ }\bibfield  {title} {\enquote {\bibinfo {title} {{Acoustic
  Streaming due to Attenuated Plane Waves}},}\ }\href {\doibase
  10.1121/1.1907010} {\bibfield  {journal} {\bibinfo  {journal} {The Journal of
  the Acoustical Society of America}\ }\textbf {\bibinfo {volume} {25}},\
  \bibinfo {pages} {68--75} (\bibinfo {year} {2005})},\ \Eprint
  {http://arxiv.org/abs/https://pubs.aip.org/asa/jasa/article-pdf/25/1/68/11472170/68\_1\_online.pdf}
  {https://pubs.aip.org/asa/jasa/article-pdf/25/1/68/11472170/68\_1\_online.pdf}
  \BibitemShut {NoStop}%
\bibitem [{\citenamefont {Xie}\ \emph {et~al.}(2002)\citenamefont {Xie},
  \citenamefont {Cao}, \citenamefont {L\"u},\ and\ \citenamefont
  {Wei}}]{Xie2002}%
  \BibitemOpen
  \bibfield  {author} {\bibinfo {author} {\bibfnamefont {W.~J.}\ \bibnamefont
  {Xie}}, \bibinfo {author} {\bibfnamefont {C.~D.}\ \bibnamefont {Cao}},
  \bibinfo {author} {\bibfnamefont {Y.~J.}\ \bibnamefont {L\"u}}, \ and\
  \bibinfo {author} {\bibfnamefont {B.}~\bibnamefont {Wei}},\ }\bibfield
  {title} {\enquote {\bibinfo {title} {Levitation of iridium and liquid mercury
  by ultrasound},}\ }\href {\doibase 10.1103/PhysRevLett.89.104304} {\bibfield
  {journal} {\bibinfo  {journal} {Phys. Rev. Lett.}\ }\textbf {\bibinfo
  {volume} {89}},\ \bibinfo {pages} {104304} (\bibinfo {year}
  {2002})}\BibitemShut {NoStop}%
\bibitem [{\citenamefont {Hirayama}\ \emph {et~al.}(2019)\citenamefont
  {Hirayama}, \citenamefont {Martinez~Plasencia}, \citenamefont {Masuda},\ and\
  \citenamefont {Subramanian}}]{Hirayama2019}%
  \BibitemOpen
  \bibfield  {author} {\bibinfo {author} {\bibfnamefont {R.}~\bibnamefont
  {Hirayama}}, \bibinfo {author} {\bibfnamefont {D.}~\bibnamefont
  {Martinez~Plasencia}}, \bibinfo {author} {\bibfnamefont {N.}~\bibnamefont
  {Masuda}}, \ and\ \bibinfo {author} {\bibfnamefont {S.}~\bibnamefont
  {Subramanian}},\ }\bibfield  {title} {\enquote {\bibinfo {title} {A
  volumetric display for visual, tactile and audio presentation using acoustic
  trapping},}\ }\href {\doibase 10.1038/s41586-019-1739-5} {\bibfield
  {journal} {\bibinfo  {journal} {Nature}\ }\textbf {\bibinfo {volume} {575}},\
  \bibinfo {pages} {320--323} (\bibinfo {year} {2019})}\BibitemShut {NoStop}%
\bibitem [{\citenamefont {Ezcurdia}\ \emph {et~al.}(2022)\citenamefont
  {Ezcurdia}, \citenamefont {Morales}, \citenamefont {Andrade},\ and\
  \citenamefont {Marzo}}]{Morales2022}%
  \BibitemOpen
  \bibfield  {author} {\bibinfo {author} {\bibfnamefont {I.~n.}\ \bibnamefont
  {Ezcurdia}}, \bibinfo {author} {\bibfnamefont {R.}~\bibnamefont {Morales}},
  \bibinfo {author} {\bibfnamefont {M.~A.~B.}\ \bibnamefont {Andrade}}, \ and\
  \bibinfo {author} {\bibfnamefont {A.}~\bibnamefont {Marzo}},\ }\bibfield
  {title} {\enquote {\bibinfo {title} {Leviprint: Contactless fabrication using
  full acoustic trapping of elongated parts.}}\ }in\ \href {\doibase
  10.1145/3528233.3530752} {\emph {\bibinfo {booktitle} {ACM SIGGRAPH 2022
  Conference Proceedings}}},\ \bibinfo {series and number} {SIGGRAPH '22}\
  (\bibinfo  {publisher} {Association for Computing Machinery},\ \bibinfo
  {address} {New York, NY, USA},\ \bibinfo {year} {2022})\BibitemShut {NoStop}%
\bibitem [{\citenamefont {Fujiwara}\ \emph {et~al.}(2011)\citenamefont
  {Fujiwara}, \citenamefont {Nakatsuma}, \citenamefont {Takahashi},\ and\
  \citenamefont {Shinoda}}]{Fujiwara2011}%
  \BibitemOpen
  \bibfield  {author} {\bibinfo {author} {\bibfnamefont {M.}~\bibnamefont
  {Fujiwara}}, \bibinfo {author} {\bibfnamefont {K.}~\bibnamefont {Nakatsuma}},
  \bibinfo {author} {\bibfnamefont {M.}~\bibnamefont {Takahashi}}, \ and\
  \bibinfo {author} {\bibfnamefont {H.}~\bibnamefont {Shinoda}},\ }\bibfield
  {title} {\enquote {\bibinfo {title} {Remote measurement of surface compliance
  distribution using ultrasound radiation pressure},}\ }in\ \href {\doibase
  10.1109/WHC.2011.5945459} {\emph {\bibinfo {booktitle} {2011 IEEE World
  Haptics Conference}}}\ (\bibinfo {year} {2011})\ pp.\ \bibinfo {pages}
  {43--47}\BibitemShut {NoStop}%
\bibitem [{\citenamefont {Hoshi}\ \emph {et~al.}(2010)\citenamefont {Hoshi},
  \citenamefont {Takahashi}, \citenamefont {Iwamoto},\ and\ \citenamefont
  {Shinoda}}]{Hoshi2010}%
  \BibitemOpen
  \bibfield  {author} {\bibinfo {author} {\bibfnamefont {T.}~\bibnamefont
  {Hoshi}}, \bibinfo {author} {\bibfnamefont {M.}~\bibnamefont {Takahashi}},
  \bibinfo {author} {\bibfnamefont {T.}~\bibnamefont {Iwamoto}}, \ and\
  \bibinfo {author} {\bibfnamefont {H.}~\bibnamefont {Shinoda}},\ }\bibfield
  {title} {\enquote {\bibinfo {title} {Noncontact tactile display based on
  radiation pressure of airborne ultrasound},}\ }\href {\doibase
  10.1109/TOH.2010.4} {\bibfield  {journal} {\bibinfo  {journal} {IEEE
  Transactions on Haptics}\ }\textbf {\bibinfo {volume} {3}},\ \bibinfo {pages}
  {155--165} (\bibinfo {year} {2010})}\BibitemShut {NoStop}%
\bibitem [{\citenamefont {Takahashi}, \citenamefont {Hasegawa},\ and\
  \citenamefont {Shinoda}(2020)}]{Takahashi2020}%
  \BibitemOpen
  \bibfield  {author} {\bibinfo {author} {\bibfnamefont {R.}~\bibnamefont
  {Takahashi}}, \bibinfo {author} {\bibfnamefont {K.}~\bibnamefont {Hasegawa}},
  \ and\ \bibinfo {author} {\bibfnamefont {H.}~\bibnamefont {Shinoda}},\
  }\bibfield  {title} {\enquote {\bibinfo {title} {Tactile stimulation by
  repetitive lateral movement of midair ultrasound focus},}\ }\href {\doibase
  10.1109/TOH.2019.2946136} {\bibfield  {journal} {\bibinfo  {journal} {IEEE
  Transactions on Haptics}\ }\textbf {\bibinfo {volume} {13}},\ \bibinfo
  {pages} {334--342} (\bibinfo {year} {2020})}\BibitemShut {NoStop}%
\bibitem [{\citenamefont {Nakajima}\ \emph {et~al.}(2021)\citenamefont
  {Nakajima}, \citenamefont {Hasegawa}, \citenamefont {Makino},\ and\
  \citenamefont {Shinoda}}]{Nakajima2021}%
  \BibitemOpen
  \bibfield  {author} {\bibinfo {author} {\bibfnamefont {M.}~\bibnamefont
  {Nakajima}}, \bibinfo {author} {\bibfnamefont {K.}~\bibnamefont {Hasegawa}},
  \bibinfo {author} {\bibfnamefont {Y.}~\bibnamefont {Makino}}, \ and\ \bibinfo
  {author} {\bibfnamefont {H.}~\bibnamefont {Shinoda}},\ }\bibfield  {title}
  {\enquote {\bibinfo {title} {Spatiotemporal pinpoint cooling sensation
  produced by ultrasound-driven mist vaporization on skin},}\ }\href {\doibase
  10.1109/TOH.2021.3086516} {\bibfield  {journal} {\bibinfo  {journal} {IEEE
  Transactions on Haptics}\ }\textbf {\bibinfo {volume} {14}},\ \bibinfo
  {pages} {874--884} (\bibinfo {year} {2021})}\BibitemShut {NoStop}%
\bibitem [{\citenamefont {Frier}\ \emph {et~al.}(2018)\citenamefont {Frier},
  \citenamefont {Ablart}, \citenamefont {Chilles}, \citenamefont {Long},
  \citenamefont {Giordano}, \citenamefont {Obrist},\ and\ \citenamefont
  {Subramanian}}]{Frier2018}%
  \BibitemOpen
  \bibfield  {author} {\bibinfo {author} {\bibfnamefont {W.}~\bibnamefont
  {Frier}}, \bibinfo {author} {\bibfnamefont {D.}~\bibnamefont {Ablart}},
  \bibinfo {author} {\bibfnamefont {J.}~\bibnamefont {Chilles}}, \bibinfo
  {author} {\bibfnamefont {B.}~\bibnamefont {Long}}, \bibinfo {author}
  {\bibfnamefont {M.}~\bibnamefont {Giordano}}, \bibinfo {author}
  {\bibfnamefont {M.}~\bibnamefont {Obrist}}, \ and\ \bibinfo {author}
  {\bibfnamefont {S.}~\bibnamefont {Subramanian}},\ }\bibfield  {title}
  {\enquote {\bibinfo {title} {Using spatiotemporal modulation to draw tactile
  patterns in mid-air},}\ }in\ \href@noop {} {\emph {\bibinfo {booktitle}
  {Haptics: Science, Technology, and Applications}}},\ \bibinfo {editor}
  {edited by\ \bibinfo {editor} {\bibfnamefont {D.}~\bibnamefont
  {Prattichizzo}}, \bibinfo {editor} {\bibfnamefont {H.}~\bibnamefont
  {Shinoda}}, \bibinfo {editor} {\bibfnamefont {H.~Z.}\ \bibnamefont {Tan}},
  \bibinfo {editor} {\bibfnamefont {E.}~\bibnamefont {Ruffaldi}}, \ and\
  \bibinfo {editor} {\bibfnamefont {A.}~\bibnamefont {Frisoli}}}\ (\bibinfo
  {publisher} {Springer International Publishing},\ \bibinfo {address} {Cham},\
  \bibinfo {year} {2018})\ pp.\ \bibinfo {pages} {270--281}\BibitemShut
  {NoStop}%
\bibitem [{\citenamefont {Hasegawa}, \citenamefont {Qiu},\ and\ \citenamefont
  {Shinoda}(2018)}]{Hasegawa2018}%
  \BibitemOpen
  \bibfield  {author} {\bibinfo {author} {\bibfnamefont {K.}~\bibnamefont
  {Hasegawa}}, \bibinfo {author} {\bibfnamefont {L.}~\bibnamefont {Qiu}}, \
  and\ \bibinfo {author} {\bibfnamefont {H.}~\bibnamefont {Shinoda}},\
  }\bibfield  {title} {\enquote {\bibinfo {title} {Midair ultrasound fragrance
  rendering},}\ }\href {\doibase 10.1109/TVCG.2018.2794118} {\bibfield
  {journal} {\bibinfo  {journal} {IEEE Transactions on Visualization and
  Computer Graphics}\ }\textbf {\bibinfo {volume} {24}},\ \bibinfo {pages}
  {1477--1485} (\bibinfo {year} {2018})}\BibitemShut {NoStop}%
\bibitem [{\citenamefont {Hasegawa}, \citenamefont {Yuki},\ and\ \citenamefont
  {Shinoda}(2019)}]{Hasegawa2019}%
  \BibitemOpen
  \bibfield  {author} {\bibinfo {author} {\bibfnamefont {K.}~\bibnamefont
  {Hasegawa}}, \bibinfo {author} {\bibfnamefont {H.}~\bibnamefont {Yuki}}, \
  and\ \bibinfo {author} {\bibfnamefont {H.}~\bibnamefont {Shinoda}},\
  }\bibfield  {title} {\enquote {\bibinfo {title} {Curved acceleration path of
  ultrasound-driven air flow},}\ }\href {\doibase 10.1063/1.5052423} {\bibfield
   {journal} {\bibinfo  {journal} {Journal of Applied Physics}\ }\textbf
  {\bibinfo {volume} {125}},\ \bibinfo {pages} {054902} (\bibinfo {year}
  {2019})},\ \Eprint {http://arxiv.org/abs/https://doi.org/10.1063/1.5052423}
  {https://doi.org/10.1063/1.5052423} \BibitemShut {NoStop}%
\bibitem [{\citenamefont {Gerchberg}(1972)}]{Gerchberg1972}%
  \BibitemOpen
  \bibfield  {author} {\bibinfo {author} {\bibfnamefont {R.~W.}\ \bibnamefont
  {Gerchberg}},\ }\bibfield  {title} {\enquote {\bibinfo {title} {A practical
  algorithm for the determination of phase from image and diffraction plane
  pictures},}\ }\href@noop {} {\bibfield  {journal} {\bibinfo  {journal}
  {Optik}\ }\textbf {\bibinfo {volume} {35}},\ \bibinfo {pages} {237--246}
  (\bibinfo {year} {1972})}\BibitemShut {NoStop}%
\bibitem [{\citenamefont {Mellin}\ and\ \citenamefont
  {Nordin}(2001)}]{Mellin2001}%
  \BibitemOpen
  \bibfield  {author} {\bibinfo {author} {\bibfnamefont {S.~D.}\ \bibnamefont
  {Mellin}}\ and\ \bibinfo {author} {\bibfnamefont {G.~P.}\ \bibnamefont
  {Nordin}},\ }\bibfield  {title} {\enquote {\bibinfo {title} {Limits of scalar
  diffraction theory and an iterative angular spectrum algorithm for finite
  aperture diffractive optical element design},}\ }\href {\doibase
  10.1364/OE.8.000705} {\bibfield  {journal} {\bibinfo  {journal} {Opt.
  Express}\ }\textbf {\bibinfo {volume} {8}},\ \bibinfo {pages} {705--722}
  (\bibinfo {year} {2001})}\BibitemShut {NoStop}%
\bibitem [{\citenamefont {Leonardo}, \citenamefont {Ianni},\ and\ \citenamefont
  {Ruocco}(2007)}]{Leonardo2007}%
  \BibitemOpen
  \bibfield  {author} {\bibinfo {author} {\bibfnamefont {R.~D.}\ \bibnamefont
  {Leonardo}}, \bibinfo {author} {\bibfnamefont {F.}~\bibnamefont {Ianni}}, \
  and\ \bibinfo {author} {\bibfnamefont {G.}~\bibnamefont {Ruocco}},\
  }\bibfield  {title} {\enquote {\bibinfo {title} {Computer generation of
  optimal holograms for optical trap arrays},}\ }\href {\doibase
  10.1364/OE.15.001913} {\bibfield  {journal} {\bibinfo  {journal} {Opt.
  Express}\ }\textbf {\bibinfo {volume} {15}},\ \bibinfo {pages} {1913--1922}
  (\bibinfo {year} {2007})}\BibitemShut {NoStop}%
\bibitem [{\citenamefont {Melde}\ \emph {et~al.}(2016)\citenamefont {Melde},
  \citenamefont {Mark}, \citenamefont {Qiu},\ and\ \citenamefont
  {Fischer}}]{Melde2016}%
  \BibitemOpen
  \bibfield  {author} {\bibinfo {author} {\bibfnamefont {K.}~\bibnamefont
  {Melde}}, \bibinfo {author} {\bibfnamefont {A.~G.}\ \bibnamefont {Mark}},
  \bibinfo {author} {\bibfnamefont {T.}~\bibnamefont {Qiu}}, \ and\ \bibinfo
  {author} {\bibfnamefont {P.}~\bibnamefont {Fischer}},\ }\bibfield  {title}
  {\enquote {\bibinfo {title} {Holograms for acoustics},}\ }\href {\doibase
  10.1038/nature19755} {\bibfield  {journal} {\bibinfo  {journal} {Nature}\
  }\textbf {\bibinfo {volume} {537}},\ \bibinfo {pages} {518--522} (\bibinfo
  {year} {2016})}\BibitemShut {NoStop}%
\bibitem [{\citenamefont {Polychronopoulos}\ and\ \citenamefont
  {Memoli}(2020)}]{Poly2020}%
  \BibitemOpen
  \bibfield  {author} {\bibinfo {author} {\bibfnamefont {S.}~\bibnamefont
  {Polychronopoulos}}\ and\ \bibinfo {author} {\bibfnamefont {G.}~\bibnamefont
  {Memoli}},\ }\bibfield  {title} {\enquote {\bibinfo {title} {Acoustic
  levitation with optimized reflective metamaterials},}\ }\href {\doibase
  10.1038/s41598-020-60978-4} {\bibfield  {journal} {\bibinfo  {journal}
  {Scientific Reports}\ }\textbf {\bibinfo {volume} {10}},\ \bibinfo {pages}
  {4254} (\bibinfo {year} {2020})}\BibitemShut {NoStop}%
\bibitem [{\citenamefont {Al~Jahdali}\ and\ \citenamefont
  {Wu}(2016)}]{Jahdali2016}%
  \BibitemOpen
  \bibfield  {author} {\bibinfo {author} {\bibfnamefont {R.}~\bibnamefont
  {Al~Jahdali}}\ and\ \bibinfo {author} {\bibfnamefont {Y.}~\bibnamefont
  {Wu}},\ }\bibfield  {title} {\enquote {\bibinfo {title} {{High transmission
  acoustic focusing by impedance-matched acoustic meta-surfaces}},}\ }\href
  {\doibase 10.1063/1.4939932} {\bibfield  {journal} {\bibinfo  {journal}
  {Applied Physics Letters}\ }\textbf {\bibinfo {volume} {108}},\ \bibinfo
  {pages} {031902} (\bibinfo {year} {2016})},\ \Eprint
  {http://arxiv.org/abs/https://pubs.aip.org/aip/apl/article-pdf/doi/10.1063/1.4939932/14475201/031902\_1\_online.pdf}
  {https://pubs.aip.org/aip/apl/article-pdf/doi/10.1063/1.4939932/14475201/031902\_1\_online.pdf}
  \BibitemShut {NoStop}%
\bibitem [{\citenamefont {Zhao}\ \emph {et~al.}(2018)\citenamefont {Zhao},
  \citenamefont {Chen}, \citenamefont {Wang},\ and\ \citenamefont
  {Zhang}}]{Zhao2018}%
  \BibitemOpen
  \bibfield  {author} {\bibinfo {author} {\bibfnamefont {S.-D.}\ \bibnamefont
  {Zhao}}, \bibinfo {author} {\bibfnamefont {A.-L.}\ \bibnamefont {Chen}},
  \bibinfo {author} {\bibfnamefont {Y.-S.}\ \bibnamefont {Wang}}, \ and\
  \bibinfo {author} {\bibfnamefont {C.}~\bibnamefont {Zhang}},\ }\bibfield
  {title} {\enquote {\bibinfo {title} {Continuously tunable acoustic
  metasurface for transmitted wavefront modulation},}\ }\href {\doibase
  10.1103/PhysRevApplied.10.054066} {\bibfield  {journal} {\bibinfo  {journal}
  {Phys. Rev. Applied}\ }\textbf {\bibinfo {volume} {10}},\ \bibinfo {pages}
  {054066} (\bibinfo {year} {2018})}\BibitemShut {NoStop}%
\bibitem [{\citenamefont {Hladky-Hennion}\ \emph {et~al.}(2013)\citenamefont
  {Hladky-Hennion}, \citenamefont {Vasseur}, \citenamefont {Haw}, \citenamefont
  {Croënne}, \citenamefont {Haumesser},\ and\ \citenamefont
  {Norris}}]{Hennion2013}%
  \BibitemOpen
  \bibfield  {author} {\bibinfo {author} {\bibfnamefont {A.-C.}\ \bibnamefont
  {Hladky-Hennion}}, \bibinfo {author} {\bibfnamefont {J.~O.}\ \bibnamefont
  {Vasseur}}, \bibinfo {author} {\bibfnamefont {G.}~\bibnamefont {Haw}},
  \bibinfo {author} {\bibfnamefont {C.}~\bibnamefont {Croënne}}, \bibinfo
  {author} {\bibfnamefont {L.}~\bibnamefont {Haumesser}}, \ and\ \bibinfo
  {author} {\bibfnamefont {A.~N.}\ \bibnamefont {Norris}},\ }\bibfield  {title}
  {\enquote {\bibinfo {title} {{Negative refraction of acoustic waves using a
  foam-like metallic structure}},}\ }\href {\doibase 10.1063/1.4801642}
  {\bibfield  {journal} {\bibinfo  {journal} {Applied Physics Letters}\
  }\textbf {\bibinfo {volume} {102}},\ \bibinfo {pages} {144103} (\bibinfo
  {year} {2013})},\ \Eprint
  {http://arxiv.org/abs/https://pubs.aip.org/aip/apl/article-pdf/doi/10.1063/1.4801642/14271362/144103\_1\_online.pdf}
  {https://pubs.aip.org/aip/apl/article-pdf/doi/10.1063/1.4801642/14271362/144103\_1\_online.pdf}
  \BibitemShut {NoStop}%
\bibitem [{\citenamefont {Zhao}\ \emph {et~al.}(2020)\citenamefont {Zhao},
  \citenamefont {Laredo}, \citenamefont {Ryan}, \citenamefont {Yazdkhasti},
  \citenamefont {Kim}, \citenamefont {Ganye}, \citenamefont {Horiuchi},\ and\
  \citenamefont {Yu}}]{Zhao2020}%
  \BibitemOpen
  \bibfield  {author} {\bibinfo {author} {\bibfnamefont {L.}~\bibnamefont
  {Zhao}}, \bibinfo {author} {\bibfnamefont {E.}~\bibnamefont {Laredo}},
  \bibinfo {author} {\bibfnamefont {O.}~\bibnamefont {Ryan}}, \bibinfo {author}
  {\bibfnamefont {A.}~\bibnamefont {Yazdkhasti}}, \bibinfo {author}
  {\bibfnamefont {H.-T.}\ \bibnamefont {Kim}}, \bibinfo {author} {\bibfnamefont
  {R.}~\bibnamefont {Ganye}}, \bibinfo {author} {\bibfnamefont
  {T.}~\bibnamefont {Horiuchi}}, \ and\ \bibinfo {author} {\bibfnamefont
  {M.}~\bibnamefont {Yu}},\ }\bibfield  {title} {\enquote {\bibinfo {title}
  {Ultrasound beam steering with flattened acoustic metamaterial luneburg
  lens},}\ }\href {\doibase 10.1063/1.5140467} {\bibfield  {journal} {\bibinfo
  {journal} {Applied Physics Letters}\ }\textbf {\bibinfo {volume} {116}},\
  \bibinfo {pages} {071902} (\bibinfo {year} {2020})},\ \Eprint
  {http://arxiv.org/abs/https://doi.org/10.1063/1.5140467}
  {https://doi.org/10.1063/1.5140467} \BibitemShut {NoStop}%
\bibitem [{\citenamefont {Memoli}\ \emph {et~al.}(2017)\citenamefont {Memoli},
  \citenamefont {Caleap}, \citenamefont {Asakawa}, \citenamefont {Sahoo},
  \citenamefont {Drinkwater},\ and\ \citenamefont {Subramanian}}]{Memoli2017}%
  \BibitemOpen
  \bibfield  {author} {\bibinfo {author} {\bibfnamefont {G.}~\bibnamefont
  {Memoli}}, \bibinfo {author} {\bibfnamefont {M.}~\bibnamefont {Caleap}},
  \bibinfo {author} {\bibfnamefont {M.}~\bibnamefont {Asakawa}}, \bibinfo
  {author} {\bibfnamefont {D.~R.}\ \bibnamefont {Sahoo}}, \bibinfo {author}
  {\bibfnamefont {B.~W.}\ \bibnamefont {Drinkwater}}, \ and\ \bibinfo {author}
  {\bibfnamefont {S.}~\bibnamefont {Subramanian}},\ }\bibfield  {title}
  {\enquote {\bibinfo {title} {Metamaterial bricks and quantization of
  meta-surfaces},}\ }\href {\doibase 10.1038/ncomms14608} {\bibfield  {journal}
  {\bibinfo  {journal} {Nature Communications}\ }\textbf {\bibinfo {volume}
  {8}},\ \bibinfo {pages} {14608} (\bibinfo {year} {2017})}\BibitemShut
  {NoStop}%
\bibitem [{\citenamefont {Suzuki}\ \emph {et~al.}(2021)\citenamefont {Suzuki},
  \citenamefont {Inoue}, \citenamefont {Fujiwara}, \citenamefont {Makino},\
  and\ \citenamefont {Shinoda}}]{autd3}%
  \BibitemOpen
  \bibfield  {author} {\bibinfo {author} {\bibfnamefont {S.}~\bibnamefont
  {Suzuki}}, \bibinfo {author} {\bibfnamefont {S.}~\bibnamefont {Inoue}},
  \bibinfo {author} {\bibfnamefont {M.}~\bibnamefont {Fujiwara}}, \bibinfo
  {author} {\bibfnamefont {Y.}~\bibnamefont {Makino}}, \ and\ \bibinfo {author}
  {\bibfnamefont {H.}~\bibnamefont {Shinoda}},\ }\bibfield  {title} {\enquote
  {\bibinfo {title} {Autd3: Scalable airborne ultrasound tactile display},}\
  }\href {\doibase 10.1109/TOH.2021.3069976} {\bibfield  {journal} {\bibinfo
  {journal} {IEEE Transactions on Haptics}\ }\textbf {\bibinfo {volume} {14}},\
  \bibinfo {pages} {740--749} (\bibinfo {year} {2021})}\BibitemShut {NoStop}%
\bibitem [{\citenamefont {Marzo}\ and\ \citenamefont
  {Drinkwater}(2019)}]{Marzo2019}%
  \BibitemOpen
  \bibfield  {author} {\bibinfo {author} {\bibfnamefont {A.}~\bibnamefont
  {Marzo}}\ and\ \bibinfo {author} {\bibfnamefont {B.~W.}\ \bibnamefont
  {Drinkwater}},\ }\bibfield  {title} {\enquote {\bibinfo {title} {Holographic
  acoustic tweezers},}\ }\href {\doibase 10.1073/pnas.1813047115} {\bibfield
  {journal} {\bibinfo  {journal} {Proceedings of the National Academy of
  Sciences}\ }\textbf {\bibinfo {volume} {116}},\ \bibinfo {pages} {84--89}
  (\bibinfo {year} {2019})},\ \Eprint
  {http://arxiv.org/abs/https://www.pnas.org/doi/pdf/10.1073/pnas.1813047115}
  {https://www.pnas.org/doi/pdf/10.1073/pnas.1813047115} \BibitemShut {NoStop}%
\bibitem [{\citenamefont {Morales}\ \emph {et~al.}(2019)\citenamefont
  {Morales}, \citenamefont {Marzo}, \citenamefont {Subramanian},\ and\
  \citenamefont {Mart\'{\i}nez}}]{Morales2019}%
  \BibitemOpen
  \bibfield  {author} {\bibinfo {author} {\bibfnamefont {R.}~\bibnamefont
  {Morales}}, \bibinfo {author} {\bibfnamefont {A.}~\bibnamefont {Marzo}},
  \bibinfo {author} {\bibfnamefont {S.}~\bibnamefont {Subramanian}}, \ and\
  \bibinfo {author} {\bibfnamefont {D.}~\bibnamefont {Mart\'{\i}nez}},\
  }\bibfield  {title} {\enquote {\bibinfo {title} {Leviprops: Animating
  levitated optimized fabric structures using holographic acoustic tweezers},}\
  }in\ \href {\doibase 10.1145/3332165.3347882} {\emph {\bibinfo {booktitle}
  {Proceedings of the 32nd Annual ACM Symposium on User Interface Software and
  Technology}}},\ \bibinfo {series and number} {UIST '19}\ (\bibinfo
  {publisher} {Association for Computing Machinery},\ \bibinfo {address} {New
  York, NY, USA},\ \bibinfo {year} {2019})\ p.\ \bibinfo {pages}
  {651–661}\BibitemShut {NoStop}%
\bibitem [{\citenamefont {Hirayama}\ \emph {et~al.}(2022)\citenamefont
  {Hirayama}, \citenamefont {Christopoulos}, \citenamefont {Plasencia},\ and\
  \citenamefont {Subramanian}}]{Hirayama2022}%
  \BibitemOpen
  \bibfield  {author} {\bibinfo {author} {\bibfnamefont {R.}~\bibnamefont
  {Hirayama}}, \bibinfo {author} {\bibfnamefont {G.}~\bibnamefont
  {Christopoulos}}, \bibinfo {author} {\bibfnamefont {D.~M.}\ \bibnamefont
  {Plasencia}}, \ and\ \bibinfo {author} {\bibfnamefont {S.}~\bibnamefont
  {Subramanian}},\ }\bibfield  {title} {\enquote {\bibinfo {title} {High-speed
  acoustic holography with arbitrary scattering objects},}\ }\href {\doibase
  10.1126/sciadv.abn7614} {\bibfield  {journal} {\bibinfo  {journal} {Science
  Advances}\ }\textbf {\bibinfo {volume} {8}},\ \bibinfo {pages} {eabn7614}
  (\bibinfo {year} {2022})},\ \Eprint
  {http://arxiv.org/abs/https://www.science.org/doi/pdf/10.1126/sciadv.abn7614}
  {https://www.science.org/doi/pdf/10.1126/sciadv.abn7614} \BibitemShut
  {NoStop}%
\bibitem [{\citenamefont {Schindel}, \citenamefont {Bashford},\ and\
  \citenamefont {Hutchins}(1997)}]{Schindel1997}%
  \BibitemOpen
  \bibfield  {author} {\bibinfo {author} {\bibfnamefont {D.}~\bibnamefont
  {Schindel}}, \bibinfo {author} {\bibfnamefont {A.}~\bibnamefont {Bashford}},
  \ and\ \bibinfo {author} {\bibfnamefont {D.}~\bibnamefont {Hutchins}},\
  }\bibfield  {title} {\enquote {\bibinfo {title} {Focussing of ultrasonic
  waves in air using a micromachined fresnel zone-plate},}\ }\href {\doibase
  https://doi.org/10.1016/S0041-624X(97)00011-5} {\bibfield  {journal}
  {\bibinfo  {journal} {Ultrasonics}\ }\textbf {\bibinfo {volume} {35}},\
  \bibinfo {pages} {275--285} (\bibinfo {year} {1997})}\BibitemShut {NoStop}%
\bibitem [{\citenamefont {Xia}\ \emph {et~al.}(2020)\citenamefont {Xia},
  \citenamefont {Li}, \citenamefont {Cai}, \citenamefont {Zhou}, \citenamefont
  {Ma},\ and\ \citenamefont {Zheng}}]{Xia2020}%
  \BibitemOpen
  \bibfield  {author} {\bibinfo {author} {\bibfnamefont {X.}~\bibnamefont
  {Xia}}, \bibinfo {author} {\bibfnamefont {Y.}~\bibnamefont {Li}}, \bibinfo
  {author} {\bibfnamefont {F.}~\bibnamefont {Cai}}, \bibinfo {author}
  {\bibfnamefont {H.}~\bibnamefont {Zhou}}, \bibinfo {author} {\bibfnamefont
  {T.}~\bibnamefont {Ma}}, \ and\ \bibinfo {author} {\bibfnamefont
  {H.}~\bibnamefont {Zheng}},\ }\bibfield  {title} {\enquote {\bibinfo {title}
  {{Ultrasonic tunable focusing by a stretchable phase-reversal Fresnel zone
  plate}},}\ }\href {\doibase 10.1063/5.0018663} {\bibfield  {journal}
  {\bibinfo  {journal} {Applied Physics Letters}\ }\textbf {\bibinfo {volume}
  {117}},\ \bibinfo {pages} {021904} (\bibinfo {year} {2020})},\ \Eprint
  {http://arxiv.org/abs/https://pubs.aip.org/aip/apl/article-pdf/doi/10.1063/5.0018663/14534672/021904\_1\_online.pdf}
  {https://pubs.aip.org/aip/apl/article-pdf/doi/10.1063/5.0018663/14534672/021904\_1\_online.pdf}
  \BibitemShut {NoStop}%
\bibitem [{\citenamefont {P{\'e}rez-L{\'o}pez}, \citenamefont {Fuster},\ and\
  \citenamefont {Candelas}(2021)}]{Perez2021}%
  \BibitemOpen
  \bibfield  {author} {\bibinfo {author} {\bibfnamefont {S.}~\bibnamefont
  {P{\'e}rez-L{\'o}pez}}, \bibinfo {author} {\bibfnamefont {J.~M.}\
  \bibnamefont {Fuster}}, \ and\ \bibinfo {author} {\bibfnamefont
  {P.}~\bibnamefont {Candelas}},\ }\bibfield  {title} {\enquote {\bibinfo
  {title} {Spatio-temporal ultrasound beam modulation to sequentially achieve
  multiple foci with a single planar monofocal lens},}\ }\href {\doibase
  10.1038/s41598-021-92849-x} {\bibfield  {journal} {\bibinfo  {journal}
  {Scientific Reports}\ }\textbf {\bibinfo {volume} {11}},\ \bibinfo {pages}
  {13458} (\bibinfo {year} {2021})}\BibitemShut {NoStop}%
\bibitem [{\citenamefont {Kitano}\ and\ \citenamefont
  {Hasegawa}(2023)}]{Kitano2023}%
  \BibitemOpen
  \bibfield  {author} {\bibinfo {author} {\bibfnamefont {M.}~\bibnamefont
  {Kitano}}\ and\ \bibinfo {author} {\bibfnamefont {K.}~\bibnamefont
  {Hasegawa}},\ }\bibfield  {title} {\enquote {\bibinfo {title} {{Airborne
  ultrasound focusing aperture with binary amplitude mask over planar
  ultrasound emissions}},}\ }\href {\doibase 10.1063/5.0140604} {\bibfield
  {journal} {\bibinfo  {journal} {Journal of Applied Physics}\ }\textbf
  {\bibinfo {volume} {133}},\ \bibinfo {pages} {144901} (\bibinfo {year}
  {2023})},\ \Eprint
  {http://arxiv.org/abs/https://pubs.aip.org/aip/jap/article-pdf/doi/10.1063/5.0140604/16824425/144901\_1\_5.0140604.pdf}
  {https://pubs.aip.org/aip/jap/article-pdf/doi/10.1063/5.0140604/16824425/144901\_1\_5.0140604.pdf}
  \BibitemShut {NoStop}%
\bibitem [{\citenamefont {Hasegawa}, \citenamefont {Fujimori},\ and\
  \citenamefont {Takasaki}(2025)}]{Hasegawa2025}%
  \BibitemOpen
  \bibfield  {author} {\bibinfo {author} {\bibfnamefont {K.}~\bibnamefont
  {Hasegawa}}, \bibinfo {author} {\bibfnamefont {M.}~\bibnamefont {Fujimori}},
  \ and\ \bibinfo {author} {\bibfnamefont {M.}~\bibnamefont {Takasaki}},\
  }\bibfield  {title} {\enquote {\bibinfo {title} {Focusing of airborne
  ultrasound emitted by a flexurally vibrating plate using a transmission mask
  with spatially designed holes (arxiv version:
  https://doi.org/10.48550/arxiv.2406.00996)},}\ }\href@noop {} {\bibfield
  {journal} {\bibinfo  {journal} {Journal of Applied Physics (in Press)}\ }
  (\bibinfo {year} {2025})}\BibitemShut {NoStop}%
\bibitem [{\citenamefont {Fink}(1992)}]{Fink1992}%
  \BibitemOpen
  \bibfield  {author} {\bibinfo {author} {\bibfnamefont {M.}~\bibnamefont
  {Fink}},\ }\bibfield  {title} {\enquote {\bibinfo {title} {Time reversal of
  ultrasonic fields. i. basic principles},}\ }\href {\doibase
  10.1109/58.156174} {\bibfield  {journal} {\bibinfo  {journal} {IEEE
  Transactions on Ultrasonics, Ferroelectrics, and Frequency Control}\ }\textbf
  {\bibinfo {volume} {39}},\ \bibinfo {pages} {555--566} (\bibinfo {year}
  {1992})}\BibitemShut {NoStop}%
\bibitem [{\citenamefont {Hasegawa}(2023)}]{Hasegawa2023}%
  \BibitemOpen
  \bibfield  {author} {\bibinfo {author} {\bibfnamefont {K.}~\bibnamefont
  {Hasegawa}},\ }\bibfield  {title} {\enquote {\bibinfo {title}
  {Two-dimensional visualization of intense ultrasound field using confined
  particle movement caused by radiation pressure field},}\ }in\ \href {\doibase
  10.1109/IUS51837.2023.10306357} {\emph {\bibinfo {booktitle} {2023 IEEE
  International Ultrasonics Symposium (IUS)}}}\ (\bibinfo {year} {2023})\ pp.\
  \bibinfo {pages} {1--4}\BibitemShut {NoStop}%
\end{thebibliography}%

\end{document}